\documentstyle[12pt]{article}
\begin{document}
\baselineskip=5.0mm
\renewcommand{\thefootnote}{\fnsymbol{footnote}}
\newcommand{\be} {\begin{equation}}
\newcommand{\ee} {\end{equation}}
\newcommand{\Be} {\begin{eqnarray}}
\newcommand{\Ee} {\end{eqnarray}}
\def\lg{\langle}
\def\rg{\rangle}
\def\a{\alpha}
\def\b{\beta}
\def\g{\gamma}
\def\G{\Gamma}
\def\d{\delta}
\def\D{\Delta}
\def\e{\epsilon}
\def\k{\kappa}
\def\l{\lambda}
\def\L{\Lambda}
\def\om{\omega}
\def\Om{\Omega}
\def\s{\sigma}
\def\Sig{\Sigma}
\def\t{\tau}
\noindent
\begin{center}
{\large
{\bf
Time-dependent optical linewidth in fluctuating environments: Stochastic
models $^*$
}}
\vspace{0.5cm}

\noindent
{\bf Gregor Diezemann} \\
\vspace{0.5cm}

\noindent
Institut f\"ur Physikalische Chemie, Universit\"at Mainz,
Welderweg 15, 55099 Mainz, FRG
\\
\end{center}
\vspace{0.5cm}
\noindent
{\it
Time-resolved optical lineshapes are calculated using a second-order
inhomogeneous cumulant expansion.
The calculation shows that in the inhomogeneous limit the optical spectra are
determined solely by two-time correlation functions.
Therefore, measurements of the Stokes-shift correlation function and the
inhomogeneous linewidth cannot provide information about the heterogeneity
lifetime for systems exhibiting dynamic heterogeneities.
The theoretical results are illustrated using a stochastic model for the
optical transition frequencies.
The model rests on the assumption that the transition frequencies are coupled
to the environmental relaxation of the system.
The latter is chosen according to a free-energy landscape model for
dynamically heterogeneous dynamics.
The model calculations show that the available experimental data are fully
compatible with a heterogeneity lifetime on the order of the primary
relaxation time.
}
\footnotetext{$^*$ dedicated to Prof. Hans Sillescu on occasion of his
65$^{\rm th}$ birthday}

\vspace{1.0cm}
\subsection*{I. Introduction}
The non-exponential response of supercooled liquids and other systems
exhibiting slow relaxation has been a topic of many investigations,
cf.\cite{vigo, EAN96}.
At present, there seems to be consencus about the fact that the primary or
$\a$-relaxation in supercooled liquids near the calorimetric glass transition
temperature $T_g$ is of a heterogeneous nature\cite{BS98, mark00}.
Several experimental techniques have been invented in order to monitor various
aspects of these dynamic heterogeneities\cite{vigo98}.
A supercooled liquid above $T_g$ represents an ergodic system on long time
scales.
Therefore, the lifetime of the dynamic heterogeneities is finite.
Up to now a couple of experimental methods exist that are able to detect
these lifetimes.
All of them are based on the idea of selecting a sub-ensemble of slowly
relaxing entities and monitor its re-equilibration afterwards.
The development of such techniques started with the invention of reduced
four-dimensional (4d) NMR\cite{4dexp}, which subsequently has been applied to
a number of glass-forming systems\cite{4dnmr, hinze98, BH98, THRS99}.
Other relevant techniques are the deep-bleach experiment designed by
Cicerone and Ediger\cite{ediger} and the atomic force microscopy method used
by Russell and Israeloff\cite{israeloff}.

A different experimental technique to monitor the dynamical behavior of
liquids is provided by time-resolved fluorescence and phosphorescence
spectroscopy, for a recent review see\cite{rankoreview}.
After an initial optical excitation with a laser-pulse, the time-dependent
emission spectra are observed, cf. Fig.1.
Most theoretical investigations of time-resolved fluorescence have
concentrated on the Stokes-shift correlation function, which is related to
the time-dependent mean $\lg\om(t)\rg$ of the emission spectrum:
\be\label{Ct.allg}
C(t)=\frac{\lg\om(t)\rg-\lg\om(\infty)\rg}{\lg\om(0)\rg-\lg\om(\infty)\rg}
\ee
In particular, the dependence of $C(t)$ on molecular variables of the solvent
have been investigated for both,
polar\cite{marcus65,ZH85,wolynes87,LYM87,HGM98}
and nonpolar solvation\cite{LS96, S397}, see also the list of references
in ref.\cite{rankoreview}.

In the field of supercooled liquids, triplet state solvation experiments have
been applied. The observed behavior of the time-resolved inhomogeneous
broadening of the optical lines has been interpreted in terms of dynamic
heterogeneities\cite{ranko97, ranko98, ranko01a}.
For one well studied system, quinoxaline (QX) in supercooled
$2$-methyltetrahydrofuran (MTHF), it has been concluded that the lifetime of
the dynamic heterogeneities or the 're-equilibration time', $\lg\t_{req}\rg$,
exceeds the mean time scale of the $\a$-relaxation, $\lg\t_\a\rg$,
considerably\cite{WR00, ranko01b}.
This is a somewhat surprising result for the following reasons.
In all 4d-NMR experiments, performed on a variety of glassformers, ranging from
polymers\cite{4dexp} to simple molecular liquids\cite{hinze98},
$\lg\t_{req}\rg\!\simeq\!\lg\t_\a\rg$ has been found. Furthermore, a strong
temperature dependence of $\lg\t_{req}\rg$ has been observed in the
deep-bleach experiments. These techniques cover the temperature range in which
the solvation experiments have been performed.
Therefore, MTHF is assumed to behave different from {\it all} systems studied
so far.
In addition, the finding of a long $\lg\t_{req}\rg$ relies on a purely
phenomenological interpretation of the data in terms of a very special
model\cite{ranko01a,ranko01b}.

It is the purpose of the present paper to present a sound theoretical
description of time-dependent optical line shapes in complex systems.
The actual calculation is based on a so-called inhomogeneous cumulant
expansion\cite{mukamel}.
The results obtained bear a formal similarity to those of ref.\cite{ranko01a},
thus partly validating the phenomological approach used there.
In order to illustrate the general results, I then treat the optical
transition frequencies as a non-Markovian stochastic process and calculate the
relevant quantities in terms of a free-energy landscape model for the primary
relaxation\cite{dieze97,DSHB98,DN99,DMO99}.
The outline of the paper is as follows. In the next section I will give a
theoretical description of optical lineshapes in the inhomogeneous limit in
terms of an inhomogeneous cumulant expansion. Some formal aspects of the
derivation are treated in Appendix A.
Sect. III gives a brief description of the free-energy landscape model
which then is used to calculate the relevant quantities.
The results are discussed in Sect. IV and the paper closes with some
conclusions.
\subsection*{II. Theory of time-dependent optical lineshapes}
In this section the theory of time-dependent optical lineshapes in the
inhomogeneous cumulant expansion will be presented.
Throughout the discussion the inhomogeneous limit will be considered
exclusively and vibronic regressions and also the finite lifetime of the
excited state will be neglected. This means, I focus on the time evolution of
the (slow, $\lg\t_\a\rg$ longer than roughly $10^{-8}s$) solvent relaxation.
To this end, the following simplified Hamiltonian of a two-state electronic
system is considered:
\be\label{Ham}
H=|g\rg[H_0+W_g]\lg g| + |e\rg[H_0+W_e+\hbar\om_{eg}]\lg e|
\ee
Here, the electronic ground and excited state are denoted by $|g\rg$ and
$|e\rg$, respectively. $H_0$ is the solvent Hamiltonian in the absence of the
solute, $W_g$ and $W_e$ denote the solute-solvent interactions in the
respective electronic states of the solute and $\hbar\om_{eg}$ is the
$0\!-\!0$ transition frequency of the solvated solute.
In the following, $\hbar$ will be set to unity.
In Appendix A the time-dependent optical lineshape is calculated
from the third order (in the electric field) response\cite{LYM87,HGM98} in
the rotating wave approximation.
In the inhomogeneous limit both, time-dependent fluorescence ($|e\rg=|S_1\rg$)
and phosphorescence ($|e\rg=|T_1\rg$) can be treated on the same footing.
In the latter case it has to be assumed additionally that intersystem crossing
and internal conversion are much faster than any other relevant time scale.
For systems with slow environmental relaxation with typical relaxation
times in the $ms\cdots s$ regime this provides no severe restriction.

In this section the optical emission lineshape will be calculated within the
framework of an inhomogeneous cumulant expansion\cite{mukamel}
starting from
\be\label{It.def}
I(\om,t)= \int_{-\infty}^\infty d\om_0P(\om,t|\om_0)S_g(\om_0)
\ee
Here, $S_g(\om)$ denotes the initial absorption spectrum, usually approximated
by the initial (non-equilibrium) emission spectrum.
Additionally, $P(\om,t|\om_0)$ is the conditional probability for finding the
transition frequency $\om$ at time $t$ given it had the value $\om_0$ at time
$t\!=\!0$.
$P(\om,t|\om_0)$ can be cast in the form
\be\label{Pt.om.int}
P(\om,t|\om_0) = \frac{1}{4\pi^2S_e(\om_0)}
		\int_{-\infty}^\infty d\t_1\int_{-\infty}^\infty d\t_2
		e^{-i\om_0\t_1} e^{i\om\t_2}
		\lg e^{i\om(0)\t_1}e^{-i\om(t)\t_2}\rg_e
\ee
which follows immediately from the definition of the joint probability,
$P(\om,t|\om_0)S_e(\om_0)=\lg\d(\om-\om(t))\d(\om_0-\om(0))\rg_e$ with
$\lg A\rg_e=Tr[A\rho_e]$, cf. Appendix A and ref.\cite{S397}.
In Appendix A, Eq.(\ref{It.def}) is derived from the general expression for
the third order response and the inhomogeneous cumulant expansion is performed
in detail, yielding Eqns.(\ref{S.om.t}, \ref{I.om.t}) for the lineshape.

Eq.(\ref{Pt.om.int}) allows to discuss the formal connection to observables
of other experimental techniques such as 2d-NMR.
In this case, the conditional probability $P(\om_L,t|\om_L')$ with $\om_L$
denoting the Larmor frequency is observed directly\cite{SRS94}.
In contrast to the optical case in NMR the dependence of $\om_L$ on molecular
orientation is known explicitly and equilibrium correlation functions
of the form $\lg e^{i\om_L(0)\t_1}e^{-i\om_L(t)\t_2}\rg$ are accessible
directly.
As is well known\cite{SRS94}, 2d-NMR spectra are not able to provide
information about dynamic heterogeneities let alone their lifetimes.
This is because in order to address this question one has to probe the
system at four points in time at least\cite{heuer97}.
It was this knowledge that led to the invention of higher-dimensional NMR
techniques\cite{4dexp}.
It is important to note that according to Eqns.(\ref{It.def},\ref{Pt.om.int})
$I(\om,t)$ does not contain any information about molecular motions beyond the
two-time correlation functions $\lg e^{i\om(0)\t_1}e^{-i\om(t)\t_2}\rg_e$ and
therefore cannot provide further insight into the issue concerning the lifetime
of dynamic heterogeneities.
Furthermore, due to the complex dependence of $\om(t)$ on all relative
distances and orientations of the solute relative to the solvent molecules,
the optical lineshapes can be interpreted in terms of molecular quantities in
a simple way only in favourable cases.
An example is the relation between the solvation coordinate and the dielectric
properties of the solvent in case of polar
solvation\cite{rankoreview,marcus65,ZH85,wolynes87,LYM87}.

Now, I proceed to perform an inhomogeneous cumulant expansion of
Eq.(\ref{Pt.om.int}).
For this purpose the Taylor expansion of the phase correlation function
$\lg e^{i\om(0)\t_1}e^{-i\om(t)\t_2}\rg_e$ is considered.
Setting $\lg\om\rg_e\!=\!0$ and $\s_e^2\!=\!\lg\om^2\rg_e$ without loss of
generality, one has in a first order approximation
\[
\lg e^{i\om(0)\t_1}e^{-i\om(t)\t_2}\rg_e\simeq
1-(\s_e^2/2)(\t_1^2+\t_2^2)+\s_e^2C(t)\t_1\t_2
\]
with the Stokes-shift correlation function given by:
\be\label{Ct.om}
C(t)=\lg\om(t)\om(0)\rg_e/\s_e^2
\ee
The non-exponential decay of $C(t)$ can conveniently be parametrized via a
Kohlrausch function, $C(t)\!=\!\exp{[(-(t/\t_K)^{\b_K})]}$.
This can be rewritten as
\be\label{Ct.lambda}
C(t)=\int\!d\l p(\l) e^{-\l t}
\ee
with relaxation rates $\l$ and a distribution function $p(\l)$.
One then has, using $\int\!d\l p(\l)\!=\!1$,
$\lg e^{i\om(0)\t_1}e^{-i\om(t)\t_2}\rg_e\simeq
\int\!d\l p(\l)\{1-(\s_e^2/2)(\t_1^2+\t_2^2)+\s_e^2e^{-\l t}\t_1\t_2\}$.
Exponentiating the expression in the curly brackets gives:
\be\label{CF.ic}
\lg e^{i\om_0(0)\t_1}e^{-i\om(t)\t_2}\rg_e\simeq
\int\!d\l p(\l)
\exp{\left\{-\frac{\s_e^2}{2}[\t_1^2+\t_2^2-2e^{-\l t}\t_1\t_2]\right\}}
\ee
It is illustrative to compare this result with the corresponding one of an
ordinary cumulant expansion. In this case one finds, without using
Eq.(\ref{Ct.lambda})
\[
\lg e^{i\om(0)\t_1}e^{-i\om(t)\t_2}\rg_e\simeq
e^{-(\s_e^2/2)[(\t_1^2+\t_2^2-2\t_1\t_2C(t)]}
\]
Therefore, this expansion does not take into account the inhomogeneities
reflected in Eq.(\ref{Ct.lambda}).
In the same way as an expansion of the form
$C(t)\!\simeq\!\int\!d\l p(\l)\!(1-\l t)\!\simeq\!\exp{[-(\int\!d\l p(\l)\l)t]}$
only yields an exponential decay the ordinary cumulant expansion can only
produce Gaussian probability distributions. The inhomogeneous cumulant
expansion instead allows one to account for a more complex
behavior\cite{mukamel}.

Performing the Fourier transform in Eq.(\ref{Pt.om.int}), using
Eq.(\ref{CF.ic}), yields for the conditional probability:
\Be\label{P.ic}
&&P(\om,t|\om_0)=\int\!d\l p(\l)P_{(\l)}(\om,t|\om_0)\nonumber\\
&&P_{(\l)}(\om,t|\om_0)=
\frac{1}{\sqrt{2\pi\s_e^2(1-e^{-2\l t})}}
    \exp{\left(-\frac{[\om-\om_0e^{-\l t}]^2}
		     {2\s_e^2(1-e^{-2\l t})}\right)}
\Ee
Note that Eq.(\ref{P.ic}) is the same as Eq.(\ref{P.cond}) for
$\s_e^2(\G)\!=\!\s_e^2$ independent of the environmental variable $\G$, which
in Eq.(\ref{P.ic}) corresponds to $\l$.
It is important to notice that both, the distribution $p(\l)$ and the
decay rates $\l$ are determined via the Stokes correlation function
$C(t)$.
In particular, the decomposition of $P(\om,t|\om_0)$ into the functions
$P_{(\l)}(\om,t|\om_0)$ is meaningful {\it only} if the $\l$ are static
quantities.
This means that the influence of possible exchange processes has to be
incorporated in the definition of these rates.
Thus, in exchange models that start from bare relaxation rates which are
modified due to exchange, the $\l$ are the effective relaxation rates and
$p(\l)$ the effective distribution.
This will be further discussed below.
Finally, the time independent absorption and emission spectra are given by:
\be\label{p0.eq}
S_g(\om)=\frac{1}{\sqrt{2\pi\s_g^2}}e^{-(\om-\D)^2/(2\s_g^2)}
\quad\mbox{and}\quad
S_e(\om)=p^{eq}(\om)=\frac{1}{\sqrt{2\pi\s_e^2}}e^{-\om^2/(2\s_e^2)}
\ee
where $\D\!=\!\lg\om\rg_g$ denoted the overall red-shift. Additionally,
$\s_g^2\!=\!\lg(\om\!-\!\D)^2\rg_g$, $\s_e^2\!=\!\lg\om^2\rg_e$ and
$\lg\om\rg_e\!=\!0$ has been used without loss of generality.
In Eq.(\ref{p0.eq}) the spectrum $S_e(\om)$ is determined by ensemble averages
with respect to the excited-state Hamiltonian, because the system occupies
this state before photon-emission.

Using the result of the inhomogeneous cumulant expansion, Eq.(\ref{P.ic}),
in Eq.(\ref{It.def}) gives a superposition of Gaussians for the optical
lineshape $I(\om,t)$.
The expectation value of any dynamical variable $A$ can then be calculated
according to
\be\label{A.mit}
\lg A(t)\rg=\int\!d\om A(\om)I(\om,t)=
\int\!d\om\int\!d\om_0 A(\om)P(\om,t|\om_0)S_g(\om_0)
\ee
From this one finds the n$^{\rm th}$ frequency moment using $A=\om^n$.
The time-dependent mean is directly related to the Stokes-shift correlation
function,
\be\label{om.t.gen}
\lg\om(t)\rg = \D\!\times\! C(t) = \D\!\times\!\int\!d\l p(\l)e^{-\l t}
\ee
and the second moment to $C(2t)$:
\be\label{om2.t.gen}
\lg\om^2(t)\rg = \s_e^2+(\D^2+\s_g^2-\s_e^2)C(2t)
\quad\mbox{where}\quad C(2t)=\int\!d\l p(\l)e^{-2\l t}
\ee
Therefore, the time-dependent variance is given by:
\be\label{sig.t.het}
\s^2(t)=\s_e^2+(\s_g^2-\s_e^2)C(t)^2+\D^2\left[C(2t)-C(t)^2\right]
\ee
This expression was found to give excellent agreement with the experimentally
observed linewidth in the triplet state solvation studies on supercooled
liquids\cite{rankoreview, ranko97, WR00}.

The results for the optical lineshape and the frequency moments obtained above
and in Appendix A can be reduced to expressions given earlier in the
literature.
If there is only one value for the environmental variable $\G$ (or $\l$)
Eq.(\ref{S.om.t}) reduces to the expression already given
above\cite{S397, kino89}.
Note that $C(t)$ does not necessarily decay exponentially in this case.
This result is obtained by performing an ordinary cumulant expansion, cf.
the discussion above.
In that case the variance is given by $\s^2(t)=\s_e^2+(\s_g^2-\s_e^2)C(t)^2$
and changes monotonously from $\s^2(0)=\s_g^2$ to $\s^2(\infty)=\s_e^2$
because of $C(0)=1$ and $C(\infty)=0$.

The result for the $P_{(\l)}(\om,t|\om_0)$ derived above, Eq.(\ref{P.ic}),
can be mapped onto an expression given by Richert\cite{ranko01a}
if his loosely defined 'local response functions' $\chi(t,\t)$ are replaced
by $e^{-\l t}$.
This point will be discussed further below.
Of course, in any theoretical calculation of a conditional probability
one can only obtain approximations to the two-time correlation function
$\lg e^{i\om(0)\t_1}e^{-i\om(t)\t_2}\rg$. In a second order cumulant
expansion the information is further reduced to the autocorrelation function
$\lg\om(t)\om(0)\rg$. As mentioned already above, no two-time correlation
function is able to provide any information about the lifetime of dynamic
heterogeneity.
\subsection*{III. A free-energy landscape model for the optical lineshape}
In order to illustrate the results derived above in this section the
transition frequencies $\om(t)$ are treated as a non-Markovian stochastic
process. For this purpose I consider the description of $\om(t)$ in terms of a
free-energy landscape model. This approach has been used previously to
describe several aspects of the $\a$-relaxation in supercooled
liquids\cite{dieze97, DSHB98}.
The basic idea is the following. The $\a$-relaxation is assumed to be
associated with activated transitions among an extensive number of metastable
glassy states or valleys in the free-energy landscape\cite{palmer82}.
This activated dynamics is described as a stochastic process $\e(t)$.
For simplicity, $\e(t)$ is modeled as a stationary Markov process.
The corresponding master equation (ME) for the conditional probability
(Green's function) $G(\e,t|\e_0)\equiv P_{1|1}(\e;t|\e_0)$ then
reads\cite{vkamp}:
\be\label{ME.e}
\frac{\partial}{\partial t}G(\e,t|\e_0)=
\int\!d\e'\k(\e|\e')G(\e',t|\e_0)
- \k(\e)G(\e,t|\e_0)\quad\mbox{with}\quad \k(\e)=\int\!d\e'\k(\e'|\e)
\ee
Here, $\k(\e|\e')$ denotes the rate for a $\e'\to \e$-transition.
The dynamics described by Eq.(\ref{ME.e}) is supposed to be relevant for the
environmental relaxation of the system.
Of course, such a model can be applied to any system exhibiting activated
dynamics, e.g. proteins, if the $\k(\e|\e')$ are chosen appropriately.
In the following calculations I choose a simple globally connected
model\cite{dieze97} for the kinetics.
After an escape out of valley (initial state $\e'$) with an activation energy
$(E_A-\e')$ any other valley (final state $\e$) can be reached with a
probability given by its density of states,
$\eta(\e)\!\propto\!e^{-\e^2/2\s(\e)^2}$:
\be\label{kap.gcm}
\k(\e|\e')=k_\infty \eta(\e)e^{\b \e'}
\ee
Here, $k_\infty$ is a prefactor and $\b\!=\!(k_{Boltzmann}T)^{-1}$.
In ref.\cite{dieze97} other models have been considered also.

The main assumption of the model is that if $x(t)$ is a process observed
experimentally, then $x(t)$ is allowed to change its value {\it solely} due
to the $\e'\to\e$ transitions.
Therefore, $x(t)$ can be considered as a 'slave process' of
$\e(t)$\cite{haken}. In particular, $x(t)$ is not Markovian.
In this context it is to be mentioned that in terms of Heuers $Q$
parameter\cite{heuer97}, which is a measure of how long a relaxation rate
remains correlated to earlier values, this model corresponds to $Q\!=\!1$
by definition, i.e. $\lg\t_{req}\rg\!=\!\lg\t_\a\rg$.
Previously, mainly the orientation of tagged particles, $x(t)=\Om(t)$, and
the position, $x(t)={\vec r}(t)$, have been considered\cite{dieze97,DSHB98}.
In the present paper, $x(t)$ is identified with the optical transition
frequencies, $x(t)=\om(t)$.
In order to describe the dynamics of $\om(t)$, different approaches can be
used.
\subsubsection*{A. Langevin equations}
A simple approach is to assign a value $\om_\e(t)$ to each of the valleys
$\e$ and consider the coupled Langevin equations:
\be\label{Lang.om}
{\dot\om}_\e(t) =
-\k(\e)\om_\e(t)+\int\!d\e'\k(\e|\e')\om_{\e'}(t) + L_{\e}(t)
\ee
where $L_\e(t)$ is a delta-correlated white noise, $\lg L_\e(t)\rg\!=\!0$,
$\lg L_\e(t)L_{\e'}(t')\rg\!=\!\G_{\e,\e'}\d(t-t')$.
For a finite number of valleys, the integral in Eq.(\ref{Lang.om}) is to be
replaced by a sum.
After diagonalization of the matrix defined in Eq.(\ref{Lang.om}) one has the
decoupled Langevin equations
\be\label{Lang.dec}
{\dot\om}_\l(t)=
-\l\om_\l(t)+L_\l(t)
\ee
where $\l$ are the eigenvalues.
These equations are directly related to Gaussian conditional
probabilities\cite{vkamp, risken}, determined by the correlation functions
$\lg\om_\l(t)\om_\l(0)\rg\!=\!\s_e^2e^{-\l t}$.
Therefore, these Langevin equations yield Eq.(\ref{P.ic}) without further
approximations.
Due to the structure of Eq.(\ref{Lang.om}) the $p(\l)$ are determined by the
$G(\e,t|\e_0)$, cf. Eq.(\ref{ME.e}). In an actual calculation, one needs
to specify how the transition frequencies $\om_\e(t)$ depend on the
environmental variables $\e$.
A very simple choice would be to neglect any correlation between $\om(t)$ and
$\e(t)$\cite{KH89}.
Without going into details of an actual calculation here, it suffices to note
that for any such  'Langevin model' the inhomogeneous cumulant expansion is
exact.
This is because in a Langevin approach one starts from Gaussian fluctuations
from the outset and $\om(t)$ represents a multivariate Ornstein-Uhlenbeck
process\cite{risken}.
This means that $\om(t)$ is a {\it non-Markovian Gaussian} process. As a
consequence, Eq.(\ref{P.ic}) results for the conditional probability.
\subsubsection*{B. Master equations}
Another approach allowing for a richer dynamics is to consider the ME
for the composite Markov process $\{\om(t),\e(t)\}$ as has been done
previously\cite{dieze97, DSHB98}.
As will be shown below, in contrast to the Langevin equations considered
above, in such an approach $\om(t)$ is {\it non-Gaussian} in general.
In the ME, the transition rates $\k(\om,\e|\om',\e')$ are chosen as simple
products, $\k(\om,\e|\om',\e')\!=\!\k(\e|\e')\L(\om|\om')$.
Here, the dimensionless functions $\L(\om|\om')$ determine the magnitude of
changes in $\om$ associated with a $\e'\!\to\!\e$-transition.
Allowing only for very small frequency changes,
$\!\om\!\to\!\om'\!=\!\om\!+\!\d$, $\d\!\ll\!1$, one can use a Kramers-Moyal
expansion of the ME as described in Appendix B.
As a result, the following equation for the conditional probability is
obtained:
\Be\label{ME.FP}
&&{\dot P}_{1|1}(\om,\e;t|\om_0,\e_0)=
-\k(\e)P_{1|1}(\om,\e;t|\om_0,\e_0)
   \nonumber\\
&&\hspace{3.9cm}
+\left[1+\rho\L^{FP}(\om)\right]
\int\!d\e'\k(\e|\e')P_{1|1}(\om,\e';t|\om_0,\e_0)
\Ee
where the Fokker Planck (FP)-operator $\L^{FP}(\om)$ is defined by:
\be\label{L.FP.def}
\L^{FP}(\om) = \frac{\partial}{\partial\om}\om
		+ \s_e^2\frac{\partial^2}{\partial\om^2}
\ee
and
\be\label{rho.def}
\rho = \frac{\d^2}{2\s_e^2}
\ee
The 'fluctuation parameter' $\rho\!\ll\!1$ measures the scale of the
$\om$-fluctuations relative to the width of the equilibrium distribution.
An analytic solution of Eq.(\ref{ME.FP}) is not feasible in general. However,
one can perform an expansion in terms of the eigenfunctions of $\L^{FP}(\om)$
which are Hermite polynomials $H_n(z)$\cite{risken}:
\be\label{P.om.e}
P_{1|1}(\om,\e;t|\om_0,\e_0) = e^{-\bar{\om}^2}
\sum_{n=0}^\infty
\frac{H_n(\bar{\om})H_n(\bar{\om}_0)}{2^nn!\sqrt{2\pi\s_\infty^2}}
G_n(\e,t|\e_0)
\quad\mbox{where}\quad \bar{\om}^2=\om^2/2\s_\infty^2
\ee
The Green's functions $G_n(\e,t|\e_0)$ are the solutions of the rate
equations:
\be\label{dt.Gn}
{\dot G}_n(\e,t|\e_0)=
-\k(\e)G_n(\e,t|\e_0)
+\left[1-n\rho\right] \int\!d\e'\k(\e|\e')G_n(\e',t|\e_0)
\ee
Note that $G_0(\e,t|\e_0)\equiv G(\e,t|\e_0)$ according to Eq.(\ref{ME.e}).
A comparison of these rate equations with those occurring in the treatment of
molecular reorientations shows that the term $n\rho$ corresponds to
$P_l(\cos{\theta})$, the $l^{\rm th}$ order Legendre polynomial of the jump
angle $\theta$. A full solution of the problem requires the calculation
of all $G_n(\e,t|\e_0)$, $n\!=\!0,1,\cdots,\infty$.
From this solution the stochastic process $\om(t)$ can be defined as a
projection of the composite Markov process $\{\om(t),\e(t)\}$.
For this purpose one has to integrate the joint probabilities over the
variables $\e$. In the special case of uncorrelated equilibrium probabilites,
$p^{eq}(\om,\e)=p^{eq}(\om)p^{eq}(\e)$, this yields the marginal
distributions\cite{dieze97}
\be\label{P.om.marg}
P(\om,t|\om_0)
=\int\!d\e_0p^{eq}(\e_0)\int\!d\e P_{1|1}(\om,\e;t|\om_0,\e_0)
\ee
From this expression and Eqns.(\ref{P.om.e}, \ref{dt.Gn}) it becomes clear
that $P(\om,t|\om_0)$ can not be written as a superposition of Gaussians.
This shows that even though Eq.(\ref{ME.FP}) resembles a Fokker-Planck
equation, $\om(t)$ is {\it not only non-Markovian but also non-Gaussian}
in general.
\subsubsection*{\small {\bf Inhomogeneous cumulant expansion}}
The non-Gaussian nature of the stochastic process $\om(t)$ in the approach
starting from a ME allows to discuss the inhomogeneous cumulant expansion
in detail.
For this purpose I proceed in exactly the same manner as in Sect.II.
There, it became clear that the only relevant quantity determining the
conditional probability is the frequency autocorrelation function.
Using the properties of the Hermite polynomials, this function can be
calculated from Eq.(\ref{P.om.e}):
\Be\label{CF.om}
\lg\om(t)\om(0)\rg=&&\hspace{-0.6cm}
\int\!d\om_0p^{eq}(\om_0)\om_0\int\!d\om\om P(\om,t|\om_0)
=\s_e^2\int\!d\e_0p^{eq}(\e_0)\int\!d\e G_1(\e,t|\e_0)\\
=&&\hspace{-0.6cm}
\s_e^2\int\!d\e_0p^{eq}(\e_0)\int\!d\e
\int\!d\l^{(1)}\Phi_{(1)}(\e,\l^{(1)})\Psi_{(1)}(\e_0,\l^{(1)})e^{-\l^{(1)}t}
\nonumber
\Ee
Here, $\Phi_{(1)}(\e,\l^{(1)})$ and $\Psi_{(1)}(\e_0,\l^{(1)})$ are the right
and left eigenvectors of the matrix defined in Eq.(\ref{dt.Gn}) for $n\!=\!1$.
The $\l^{(1)}$ are the corresponding eigenvalues.
Performing the same calculation as in Sect.II, one recovers Eq.(\ref{P.ic})
for the conditional probability, $P(\om,t|\om_0)$, with
\be\label{P.om}
p(\l^{(1)})=\int\!d\e_0p^{eq}(\e_0)\int\!d\e
	      \Phi_{(1)}(\e,\l^{(1)})\Psi_{(1)}(\e_0,\l^{(1)})
\ee
Thus, the inhomogeneous cumulant expansion provides an explicit expression
for the 'distribution of relaxation rates', $p(\l)$.

Note, that a Gaussian approximation is equivalent to approximating the
Green's functions $G_n(\e,t|\e_0)$, $n\!>\!1$, by $G_n(\e,t|\e_0)\!\simeq\!
\int\!d\l_1\Phi_{(1)}(\e,\l^{(1)})\Psi_{(1)}(\e_0,\l^{(1)})e^{-n\l^{(1)} t}$.
Inserting this expression into Eq.(\ref{P.om.e}) allows one to compute the sum
over $n$ resulting in a Gaussian\cite{risken}. Using Eq.(\ref{P.om.marg})
yields Eq.(\ref{P.ic}) with $p(\l)$ given by Eq.(\ref{P.om}).
It is important to point out that the frequency autocorrelation function,
Eq.(\ref{CF.om}), also determines the conditional probability in an
ordinary cumulant expansion, cf. the discussion in Sect.II.
Therefore, in the present context Eq.(\ref{P.ic}) is the result of an
{\it inhomogeneous Gaussian approximation}.

In general the assumption of Gaussian fluctuations presents an approximation,
only. There are, however, two limits in which $\om(t)$ becomes a Gaussian
stochastic process. As shown in Appendix C a perturbation expansion of
Eq.(\ref{dt.Gn}) yields a Gaussian probability distribution in the limits
of short and long times, respectively.

In the short time limit, one has to replace the rates $\l$ in Eq.(\ref{P.ic})
by $\rho\k(\e)$ and to set $p(\l)\!=\!p^{eq}(\e)$.
The relaxation rates are therefore determined by the escape rates $\k(\e)$.
This means that any single $\e\!\to\!\e'$ transition is effective for the
relaxation of the transition frequencies $\om(t)$.
This limit therefore corresponds to a situation in which the $\om(t)$
fluctuate in a quasi-static environment.

In the long time limit, only the average relaxation rate,
$\rho\lg\k\rg=\int\!d\e\k(\e)$, determines the temporal behavior of
$P(\om,t|\om_0)$ and one has to set $p(\l)\!=\!\d(\l\!-\!\rho\lg\k\rg)$ in
Eq.(\ref{P.ic}).
Therefore, in this limit $\om(t)$ is an Ornstein-Uhlenbeck process\cite{vkamp}
which means that $\om(t)$ is Markovian and Gaussian in this limit. Only the
time scale for relaxation is set by the mean environmental relaxation rate
$\lg\k\rg$.

The inhomogeneous Gaussian approximation discussed above smoothly
interpolates between these two limits, if $\rho\!\ll\!1$. Here, the
difference between an inhomogeneous cumulant expansion and an ordinary one
can be seen most clearly. In both cases the correct long time limit is
approached, as is well known from the theory of random walks\cite{vkamp}.
The correct short time limit, however, is only reached if the inhomogeneous
cumulant expansion is performed.
\subsubsection*{\small {\bf Model calculations}}
In Fig.2 the Stokes-shift correlation function $C(t)$, calculated according to
Eq.(\ref{CF.om}), and the time-dependent part of the variance,
Eq.(\ref{sig.t.het}), are shown for $\rho\!=\!0.1$ and $\s_g\!=\!\s_e$.
The parameters, $\k_\infty$ and $\s(\e)$, cf. Eq.(\ref{kap.gcm}),
are chosen in such a way that $C(t)$ is well parametrized by a Kohlrausch
function, $C(t)=\exp{[-(t/\t_K)^{\b_K}]}$.
For the curves with $\b_K=0.5$, the choice was guided by the experimental
results of Wendt and Richert on triplet state solvation in MTHF\cite{WR00}.
As already mentioned in the last section, the shown behavior for
$C(2t)-C(t)^2=[\s^2(t)-\s_e^2]/\D^2$ is in excellent agreement with the
experimental data. Comparing the different curves in Fig.2, it is seen that
for more stretched $C(t)$ the range of positive  $C(2t)-C(t)^2$ is increased.
Also the absolute value increases with decreasing $\b_K$.
Without showing the results of calculations for other choices of the
fluctuation parameter $\rho$ here I just mention that this has only a small
effect on the overall behavior if $\rho$ is not chosen too large.
In this latter case, however, the Kramers-Moyal expansion is no longer
meaningful and Eq.(\ref{ME.FP}) ceases to be valid.

As has been pointed out, the solution of Eq.(\ref{ME.FP}) results in a
stochastic process $\om(t)$ that is not only non-Markovian, but also
non-Gaussian.
A detailed study of the deviations from Gaussian behavior of the exact
solution of Eq.(\ref{ME.FP}) is difficult in general.
However, one can compare the low-order moments calculated in the inhomogeneous
Gaussian approximation, Eqns.(\ref{P.ic},\ref{P.om}), with those obtained using
the exact Eqns.(\ref{P.om.e},\ref{dt.Gn}).
In a way similar to the calculation of $\lg\om(t)\om(0)\rg$ one finds
for $\s_e\!=\!\s_g$
\be\label{Moms.ng}
\lg\om(t)\rg_{(NG)}=\D\!\times\! C(t)
\quad\mbox{and}\quad
\lg\om^2(t)\rg_{(NG)}=
\s_e^2+\D^2\int\!d\e p^{eq}(\e)\!\int\!d\e_0 G_2(\e,t|\e_0)
\ee
The first moment of course coincides with the Gaussian approximation,
Eq.(\ref{om.t.gen}), $\lg\om(t)\rg_{(NG)}\equiv\lg\om(t)\rg$.
For the second moment and the variance, $\s^2(t)$, this is not
true. Here, Eq.(\ref{sig.t.het}) has to be compared to
\be\label{sig.t.ng}
\s^2(t)_{(NG)}=\s_e^2+\D^2\xi(t)
\quad\mbox{where}\quad
\xi(t)=\int\!d\e p^{eq}(\e)\!\int\!d\e_0 G_2(\e,t|\e_0)-C(t)^2
\ee
This quantity is plotted in Fig.3(a) for the same parameters as used in Fig.2
for $\b_K=0.5$ along with $C(2t)-C(t)^2$ for comparison.
It is clearly seen that $\xi(t)$ exceeds $C(2t)-C(t)^2$ in the whole
range where both functions are nonzero.
In order to qualitatively discuss higher order moments $\lg\om^n(t)\rg_{(NG)}$
it is sufficient to notice that these are determined by
$\int\!d\e p^{eq}(\e)\!\int\!d\e_0 G_n(\e,t|\e_0)$, as can be seen from
Eq.(\ref{P.om.e}).
Therefore, the functions
\be\label{zeta.n}
\zeta_n(t)=\int\!d\e p^{eq}(\e)\!\int\!d\e_0 G_n(\e,t|\e_0)-C(nt)
\ee
give a qualitative measure for the deviations of $\lg\om^n(t)\rg_{(NG)}$ from
a Gaussian behavior. This is because in the Gaussian approximation
the $\zeta_n(t)$ vanish identically, $\zeta_n(t)\!\equiv\!0$.
The $\zeta_n(t)$ are plotted in Fig.3(b) for $n\!=\!2,4,6,8$.
From this plot it is evident that the deviations from Gaussian behavior even
increase with $n$.
Thus, if one wants to go beyond the Gaussian approximation, it is not
sufficient to consider only the first few moments. In particular, the second
moment gives only a poor approximation to the inhomogeneous linewidth.
Additionally, the Stokes-shift correlation function is no longer determined
by the first moment but also higher moments become important.
However, it should be kept in mind that if higher order terms are kept in the
cumulant expansion of $P(\om,t|\om_0)$, these terms also have to be taken into
account for the static spectra $S_e(\om)$ for reasons of internal consistency,
cf. the discussion in Appendix A.
As will be discussed further below, it is the identification of
$\s^2(t)_{(NG)}$ with the inhomogeneous linewidth for one special exchange
model that let Richert\cite{ranko01b} to the incorrect conclusion that models
with $\lg\t_{req}\rg\!\simeq\!\lg\t_\a\rg$ are incompatible with the
experimental data obtained for QX in MTHF\cite{WR00}.
\subsection*{IV. Discussion}
In Sect.II and Appendix A an inhomogeneous cumulant expansion has been
performed to calculate the time-dependent optical linewidths in complex
systems like supercooled liquids or proteins.
Because the expansion has been truncated at second order, the transition
frequencies $\om(t)$ constitute a stochastic process that is {\it Gaussian} but
{\it non-Markovian} in this approximation.
In Sect.III the theoretical results have been illustrated by a specific
stochastic model for $\om(t)$.
The coupling of these frequencies to the environmental fluctuations, treated
in terms of the free-energy landscape model, renders $\om(t)$ non-Markovian.
The same is true if exchange models are considered instead\cite{DHS01}.

In theoretical treatments of polar solvation\cite{ZH85,wolynes87,LYM87} a
simple interpretation of the time-dependent Stokes shift $C(t)$ and
accordingly the solvation coordinate is given in terms of the dielectric
properties of the solvent.
If the solvation coordinate can be modeled as a Gaussian Markov process,
also the interpretation in terms of solvation free energies is
straightforward\cite{ZH85,LYM87}.
This, however, usually leads to exponential relaxation.
If instead the Stokes-shift correlation function decays non-exponentially,
such a simple relation is no longer obvious as in this case the solvation
coordinate does not constitute a Markov process.
Rewriting the non-exponentially decaying Stokes-shift correlation
function as in Eq.(\ref{Ct.lambda}) still allows to relate $C(t)$ to the
dielectric response via an expression of the form
$C(t)\sim\int\!d\l p(\l)\exp{(-t/\t_{diel}(\l))}$.
Here, $\t_{diel}(\l)$ denote dielectric relaxation times, the values of
which depend on the model considered\cite{rankoreview}.
Thus, the treatment of $\om(t)$ as a projection from a higher-dimensional
Markov process along with the inhomogeneous cumulant expansion might be helpful
in the interpretation of the solvation coordinate in more complex situations.

The free-energy landscape model for the dynamics used in the last section
to treat the spectral diffusion process is of a phenomenological nature and
specific assumptions about the dynamical evolution of the 'order parameter'
$\e(t)$ and the coupling between the different processes have to be made in
order to obtain a tractable model.
One of the main features of this model lies in the fact that there is no
difference between 'relaxation rates' and 'exchange rates' as in typical
exchange models\cite{sill96}. Additionally, as mentioned above, the choice
of the $\k(\e|\e')$ according to Eq.(\ref{kap.gcm}) corresponds to a minimum
rate memory, $Q\!\equiv\!1$\cite{heuer97}. This means that the the lifetime of
the dynamic heterogeneities is the same as the time scale of the
$\a$-relaxation.

The treatment of the spectral diffusion process in terms of coupled
Langevin equations, Eq.(\ref{Lang.om}), immediately gives the result of
the inhomogeneous cumulant expansion, Eq.(\ref{P.ic}), for the conditional
probability. Consequently, the first moment is given by Eq.(\ref{om.t.gen})
and the time-dependent variance by Eq.(\ref{sig.t.het}).
The reason for this behavior is given by the fact that in such a Langevin
equation model one starts from Gaussian fluctuations from the outset.
Therefore, $\om(t)$ represents a non-Markovian Gaussian process.
According to ref.\cite{WR00} this means that this model gives excellent
agreement with the experimental data on QX in MTHF if the transition rates
are chosen in such a way that the Stokes-shift correlation is well
parameterized by a Kohlrausch function with $\b_K\!=\!0.5$.
Therefore, the experimental data are fully compatible with $Q\!\equiv\!1$.

A richer dynamical behavior of $\om(t)$ can be studied by considering the more
general case of a ME for the composite Markov process $\{\om(t),\e(t)\}$.
The Kramers-Moyal expansion of this ME has led to the introduction of the
parameter $\rho=(\d^2/2\s_e^2)$ that measures the fluctuations in transition
frequency relative to width of the equilibrium steady state emission spectrum.
It has already been mentioned in the context of Eq.(\ref{dt.Gn}) that the
fluctuation parameter $\rho$ plays a role very similar to the jump angle in
models for the reorientational motion.
2d-NMR experiments performed in the time domain allow to extract these jump
angles, roughly on the order of $10^o$ in supercooled
liquids\cite{hinze98, BH98, burki98}.
As it is the reorientational and translational motion of the solvent molecules
that is responsible for the $\om$-fluctuations, one expects that $\rho$ also
takes on finite values.
Note that such finite fluctuations cannot be treated in terms of Langevin
equations.

In general, the process $\om(t)$ is neither Markovian {\it nor} Gaussian in
the treatment using the ME.
Only in two limiting situations the result of the inhomogeneous cumulant
expansion, Eq.(\ref{P.ic}), is found for the conditional probability for small
$\rho$.
At short times the transition frequencies fluctuate in a quasi-static
environment and every $\e\!\to\!\e'$ transition gives rise to relaxation.
Therefore, in this limit the relevant rates are just the 'escape rates'
$\k(\e)\!=\!\int\!d\e'\k(\e'|\e)$.
Conversely, at long times the environmental relaxation appears to be fast
compared to the changes in $\om(t)$. Thus, only the average rate
$\lg\k\rg$ is relevant for the time scale of the $\om$-fluctuations,
rendering $\om(t)$ an Ornstein-Uhlenbeck process.
Note that these two limits can be compared to the so-called
short-time approximation and long-time approximation in stochastic
lineshape theory\cite{kuboII}.
Apart from these limits, $\om(t)$ is not a Gaussian process and the deviations
from Gaussian behavior have been discussed in context of Fig.3.
Furthermore, I have discussed an inhomogeneous Gaussian approximation for the
conditional probability which allows to smoothly interpolate between the
limits of short and long times, where it becomes exact.
For this approximation, the result of the inhomogeneous cumulant expansion,
Eq.(\ref{P.ic}), is recovered.
In contrast, an ordinary cumulant expansion fails to reproduce the exact
behavior at short times.

It has been mentioned already that the expression for the
$P_{(\l)}(\om,t|\om_0)$, Eq.(\ref{P.ic}), can be compared to a corresponding
one given by Richert\cite{ranko01a}, if his $\chi(t,\t)$ are replaced by
$\exp{(-\l t)}$.
However, in ref.\cite{ranko01a} these functions were denoted as conditional
probabilities and an expression similar to Eq.(\ref{P.ic}) was {\it postulated}
to hold rather than derived from a theoretical treatment.
Additionally, the meaning of the 'local response function' $\chi(t,\t)$
remains unclear.
Unfortunately, no definition for these objects was given apart from the fact
that they are related to the Stokes-shift correlation function via
$C(t)=\lg\chi(t,\t)\rg$ which contradicts the term 'response function'.
Thus, it appears that a meaningful interpretation of the $\chi(t,\t)$
can be given only if these functions are identified with $\exp{(-\l t)}$.
Surprisingly, this apparently is not Richert's interpretation because he
concludes that the time-dependent variance is determined by the
'spatially distributed local solvent responses' $\chi(t,\t)$ in case of a
static heterogeneity. Additionally, in ref.\cite{ranko01b} he
considers exchange models in which the 'time constants $\t$ of the local
responses' fluctuate in time.
He furthermore claims that the experimental results are only compatible with
slow rate exchange.
Of course, in an inhomogeneous cumulant expansion the rates $\l$ (or
corresponding time constants $\t$) have to be static quantities.
As I have shown in the last section, stochastic models which do not
distinguish at all between 'relaxation' and 'exchange' give results
compatible with the available experimental data\cite{WR00}.
The same holds for exchange models\cite{DHS01}.
The confusion seems to have its origin in the fact that the exchange models
studied in ref.\cite{ranko01b} are not Gaussian.
Therefore, a Gaussian approximation has to be applied to the models in order
to be able to compare the results to an inhomogeneous cumulant expansion
truncated at the second order.
Otherwise, $C(t)$ is not determined by the first moment and the linewidth
is not given by $\s^2(t)$.
As noted above, including higher order terms is non-trivial, as this should
also be done for the static spectra.
These difficulties can be avoided if one considers models which are
Gaussian intrinsically, like Langevin equation models.
\subsection*{IV. Conclusions}
In the present paper I have derived an expression for the time-dependent
optical linewidths in the inhomogeneous limit for systems exhibiting slow
relaxation. For this purpose, an inhomogeneous cumulant expansion has been
performed, utilizing the fact that a non-exponentially decaying Stokes-shift
correlation function can be represented as a weighted sum of exponentially
decaying functions.

In a free-energy landscape model developed to model the primary relaxation in
supercooled liquids, the optical transition frequencies were treated as a
stochastic process $\om(t)$, which is not Markovian in general.
Two variants of this model have been considered, a Langevin approach and
a treatment in terms of a master equation.
In the former case the model is solved exactly by the inhomogeneous
cumulant expansion as the fluctuations truly are Gaussian.
Therefore, in this case $\om(t)$ is a non-Markovian Gaussian process.
This does not hold for the more general ME approach.
Here, the conditional probabilities are Gaussians only in the short time
and the long times limits. I have additionally considered an inhomogeneous
Gaussian approximation that interpolates between these limits.
In general, however, in such more complex models $\om(t)$ is not only
non-Markovian but additionally non-Gaussian.

Due to the fact that the system is probed at two times only in the optical
experiments under consideration, one cannot obtain any information about
the lifetime of dynamic heterogeneities from these experiments
alone\cite{heuer97}.
This is in contrast to recent claims where it was concluded on the existence
of long lived heterogeneities within a special model scenario that was
{\it assumed} to apply\cite{ranko01b}.
The model calculations performed in the present paper clearly show that the
experimental data are also fully compatible with the assumption that the
lifetime of the dynamic heterogeneities is on the same order as the
$\a$-relaxation time.
\subsubsection*{Acknowledgement}
I am grateful to Hans Sillescu for illuminating discussions and his continuous
encouragement over the years. I thank Roland B\"ohmer and Gerald Hinze for fruitful
discussions. Financial support of our work on dynamic heterogeneities from the
Deutsche Forschungsgemeinschaft via the Sonderforschungsbereich 262 is
acknowledged.
\newpage
\begin{appendix}
\subsection*{Appendix A: Nonlinear response theory of optical lineshapes}
\setcounter{equation}{0}
\renewcommand{\theequation}{A.\arabic{equation}}
In this Appendix the result for the time-dependent optical lineshapes will be
derived from an inhomogeneous cumulant expansion of the corresponding
third order optical response.

The starting point is the Hamiltonian, Eq.(\ref{Ham}), to which the
radiation-matter interaction responsible for the electronic transitions is
added. In the Condon approximation the latter is simply given by
$H_{int}=|g\rg\lg e|+|e\rg\lg g|$.
Treating the excitation pulse classically and the emitted radiation quantum
mechanically, the following expression for the time-resolved lineshape is
obtained in the rotating wave approximation, see eg.\cite{LYM87,HGM98}:
\Be\label{St.allg}
S(\om_S,\om_L,t)\propto
&&\hspace{-0.6cm}
Re\int_{-\infty}^\infty\!d\t_3\int_{-\infty}^\infty\!d\t_2
\int_{-\infty}^\infty\!d\t_1e^{i(\om_Lt_1+\om_St_3)}\nonumber\\
&&\hspace{-0.6cm}
E_L(t-t_1-t_2-t_3)E_L^*(t-t_2-t_3) R(t_1,t_2,t_3)
\Ee
where $E_L(t)$ denotes the time-dependent amplitude of the incoming field
and $R(t_1,t_2,t_3)$ is a third order response function which in the inhomogeneous
limit is given by\cite{LYM87}:
\be\label{R.def}
R(t_1,t_2,t_3) = Tr\left[e^{-iUt_3}e^{-i{\cal L}_et_2}e^{iUt_1}\varrho_g\right]
\ee
In this expression, $U\!=\!W_e\!-\!W_g$ and $\varrho_g$ denotes the canonical
ground state density operator, $\varrho_g=\exp{(-\b H_g)}/Tr[\exp{(-\b H_g)}]$
with $H_g=\lg g|H|g\rg$ and $\b=(k_BT)^{-1}$.
Furthermore, ${\cal L}_e$ denotes the Liouville operator corresponding to
$H_e$\cite{mukamel}.


Eq.(\ref{St.allg}) describes the processes visualized in the double-sided
Feynman diagram\cite{mukamel} in Fig.4.
It is to be noted that a term corresponding to a resonant Raman process has
been neglected completely in Eq.(\ref{St.allg}).

Following ref.\cite{LYM87}, in a next step it is assumed that the solvation
coordinate $U$ is the only relevant quantity determing the solvent relaxation.
After a projection onto the operator $U$ and a second order cumulant
expansion, one gets a simple expression for the response function
$R(t_1,t_2,t_3)$, which is determined solely by the Stokes-shift correlation
function
$Tr\left[\{e^{iH_et}Ue^{-iH_et}\}U_e(0)\varrho_e\right]$ and some static
quantities.
Therefore, in the inhomogeneous limit the optical lineshape is completely
determined by a two-time correlation function.

In order to perform an inhomogeneous cumulant expansion\cite{mukamel},
it is assumed that all quantities depend on a configurational
variable $\G$ chosen from some distribution $p(\G)$.
One then calculates
$R(t_1,t_2,t_3)=\int\!d\G p(\G) R(t_1,t_2,t_3;\G)$, cf.\cite{mukamel},
with a result analogously to that obtained in ref.\cite{LYM87}:
\Be\label{R.inh}
R(t_1,t_2,t_3)=
&&\hspace{-0.6cm}
\int\!d\G p(\G) \exp{(im_e(\G)t_1)}
\exp{(it_3[m_e(\G)+(m_g(\G)-m_e(\G))C(t_2;\G)]}\times\nonumber\\
&&\hspace{0.65cm}
\times\exp{(-t_3^2/2[\s_e^2(\G)+(\s_g^2(\G)-\s_e^2(\G))C^2(t_2;\G)])}\times\\
&&\hspace{0.65cm}
\times\exp{(-t_1^2/2\s_g^2(\G)-t_1t_3\s_g^2(\G)C^2(t_2;\G))}\nonumber
\Ee
where
\Be\label{Mom.def}
&&m_\a(\G)=\lg\om(0;\G)\rg_\a\quad,\quad
\s_\a^2(\G)=\lg (\om^2(0;\G)\rg_\a-m_\a^2(\G)\nonumber\\
&&C(t;\G)=(\lg (\om(t;\G)\om(0;\G)\rg_e-m_e^2(\G))/\s_e^2(\G)
\Ee
Here, $\lg A(t;\G)\rg_\a\!=\!Tr\left[A(t;\G)\varrho_\a\right]$, $\a\!\in\!\{e,g\}$
and $\om(t;\G)$ is the classical analogue of $(\om_{eg}(\G)+U(t;\G))$.
As in Eq.(\ref{R.inh}) one has
$\lg A(t)\rg_\a=\int\!d\G p(\G)\lg A(t;\G)\rg_\a$.

A simple expression for $S(\om_S,\om_L,t)$ is obtained by assuming that
$t_1$ is very short and additionally neglecting the optical coherence during
$t_3$. In this case $t_2$ can be replaced with the observation time $t$,
cf. Fig.4.
Formally, these approximations amount to replace
$E_L(t-t_1-t_2-t_3)E_L^*(t-t_2-t_3)$ in Eq.(\ref{St.allg}) by
$|E_L|^2\d(t_2-t)\d(t_1)$. The remaining $t_3$-integration can be performed
and one finds for the $\om_L$-independent normalized lineshape:
\be\label{S.om.t}
I(\om_S,t)=\int\!d\G p(\G)\frac{1}{\sqrt{2\pi\Sig^2(t;\G)}}
 \exp{\left(-\frac{\left\{\om_S-[m_e(\G)+(m_g(\G)-m_e(\G))C(t;\G)]\right\}^2}
		     {2\Sig^2(t;\G)}\right)}
\ee
with
\be\label{Sig.t}
\Sig^2(t;\G)=\s_e^2(\G)+(\s_g^2(\G)-\s_e^2(\G))C^2(t;\G)
\ee
Alternatively, Eq.(\ref{S.om.t}) can be cast in the form
\be\label{I.om.t}
I(\om,t)=\int\!d\G p(\G) \int\!d\om' P_{(\G)}(\om,t|\om')S_g(\om';\G)
\ee
where $S_\a(\om;\G)=
[2\pi\s_\a^2(\G)]^{-1/2}\exp{[(-(\om-m_\a(\G))^2/(2\s_\a^2(\G))])}$.
$P_{(\G)}(\om,t|\om')$ is the conditional probability to find the
emission frequency $\om$ at time $t$, given that it had the value
$\om'$ at time $t=0$,
\be\label{P.cond}
P_{(\G)}(\om,t|\om_0)=\frac{1}{\sqrt{2\pi\s_e^2(\G)(1-C^2(t;\G))}}
    \exp{\left(-\frac{[\om-\om_0C(t;\G)]^2}
		     {2\s_e^2(\G)(1-C^2(t;\G))}\right)}
\ee
From the properties of the conditional probability\cite{vkamp} one has
\be\label{I.om.grenz}
I(\om,0)=\int\!d\G p(\G) S_g(\om;\G)
\quad\mbox{and}\quad
I(\om,\infty)=\int\!d\G p(\G) S_e(\om;\G)
\ee
One might argue that the derivation given here is rather formal,
especially as the result again can be cast into a form that resembles ordinary
linear response theory, Eq.(\ref{I.om.t}).
The reason for the formal treatment is twofold.
From the inhomogeneous cumulant expansion for the response function,
Eq.(\ref{R.inh}), it becomes clear that the 'ingredients' in Eq.(\ref{I.om.t})
are not independent.
This does not become clear from Eq.(\ref{I.om.t}) alone.
If only this expression would be given, there would be no reason for
restricting the cumulant expansions for $S_g(\om';\G)$ and the conditional
probability to the same order.
Also, a generalization of the expression
$I(\om,t)= \int_{-\infty}^\infty d\om_0P(\om,t|\om_0)S_g(\om_0)$, cf.
ref.\cite{S397}, is not straightforward in general. In particular it is not
clear how to perform the inhomogeneous cumulant expansion. Only in the limit
of $\G$-independent static quantities, it suffices to consider
$P(\om,t|\om_0)$.

In the text, only the case of $\G$-independent static quantities
$m_\a$ and $\s_\a$ is considered for simplicity.
This means that one has $S_\a(\om)=S_\a(\om;\G)$ independent of $\G$.
Additionally, this allows to rewrite Eq.(\ref{I.om.t}) as
$I(\om,t)= \int_{-\infty}^\infty d\om_0P(\om,t|\om_0)S_g(\om_0)$.
In the text I choose $m_e=0$ and additionally rename $m_g=\D$ denoting the
overall red-shift.
%
\subsection*{Appendix B: Kramers Moyal expansion}
\setcounter{equation}{0}
\renewcommand{\theequation}{B.\arabic{equation}}
In this appendix the derivation of Eq.(\ref{ME.FP}) is outlined.
The starting point is the ME
\Be\label{ME.x.om.allg}
\frac{\partial}{\partial t}P_{1|1}(\om,\e;t|\om_0,\e_0)
&&\hspace{-0.6cm}
=\int\!d\om'\L(\om|\om')\int\!d\e'\k(\e|\e')P_{1|1}(\om',\e';t|\om_0,\e_0)\\
&&\hspace{-0.6cm}
-\int\!d\om'\L(\om'|\om)\int\!d\e'\k(\e'|\e)P_{1|1}(\om,\e;t|\om_0,\e_0)\nonumber
\Ee
As noted in the text, the $\L(\om|\om')$ are assumed to be finite for small
jump-lengths only. This means, that $\om(t)$ is modelled as a so-called
one-step process\cite{vkamp, gardiner} with a jump length $\d$
($\d^2\!=\!2\s_e^2\rho$, cf. Eq.(\ref{rho.def})). Furthermore, it is assumed
that spectral diffusion in the limit of vanishing $\d$ can be viewed as a
diffusion in a potential of the form $U(\om)=(4\b\s_e^2)^{-1}\om^2$
with $\b=(k_BT)^{-1}$. (The factor $1/4$ is introduced here for
convenience only.)
The force originating from this potential ensures that the time-dependent
distribution approaches the equilibrium distribution $p^{eq}(\om)$,
Eq.(\ref{p0.eq}), in the long time limit.
With the functions $A(\om)=-\rho\om$ and $B(\om)=2\rho\s_e^2$ the
$\L(\om|\om')$ can be written in the from
\be\label{L.KM.om}
\L(\om|\om') =
 \left[
      \frac{A(\om)}{2\d} + \frac{B(\om)}{2\d^2}
 \right]\d(\om'-[\om+\d])
+ \left[
      -\frac{A(\om)}{2\d} + \frac{B(\om)}{2\d^2}
 \right]\d(\om'-[\om-\d])
\ee
The occurence of the inverse powers of $\d$ has its origin in the scaling
properties of the $\L(\om|\om')$\cite{gardiner}.
In the limit of small $\d$ a Kramers-Moyal (KM) expansion (or equivalently in
this case van Kampen's system size expansion\cite{vkamp}) can be performed.
With the definition of the jump moments
$\a_n(\om)=\int\!d\om'(\om'-\om)^n\L(\om'|\om)$\cite{vkamp} and the relations
\[
\int\!d\om'\L(\om|\om')f(\om')=
\sum_{n=0}^\infty \frac{(-1)^n}{n!}
\left(\frac{\partial}{\partial\om}\right)^n \a_n(\om)f(\om)
\!\!\quad\mbox{and}\quad\!\!
\int\!d\om'\L(\om|\om')f(\om)=\a_0(\om)f(\om)
\]
holding for an arbitrary function $f(\om)$ one finds
\Be\label{ME.KM.allg}
&&\frac{\partial}{\partial t}P_{1|1}(\om,\e;t|\om_0,\e_0)=
-\a_0(\om)\k(\e)P_{1|1}(\om,\e;t|\om_0,\e_0)
   \nonumber\\
&&\hspace{3.9cm}
+\left[\a_0(\om)+\L^{KM}(\om)\right]
\int\!d\e'\k(\e|\e')P_{1|1}(\om,\e';t|\om_0,\e_0)
\Ee
with $\L^{KM}(\om)=\sum_{n=1}^\infty \frac{(-1)^n}{n!}
\left(\frac{\partial}{\partial\om}\right)^n \a_n(\om)$.
Using $\a_0(\om)=1$, $\a_1(\om)=-\rho\om$ and $\a_2(\om)=\rho2\s_e^2$
in Eq.(\ref{ME.KM.allg}) and truncating the expansion at the second order
directly yields Eq.(\ref{ME.FP}) in the text.
The validity of the truncation is controlled by the smallness of the jump
moments $\a_3(\om)=-2\rho^2\s_e^2\om$ and $\a_4(\om)=4\s_e^4\rho^2$.
%
\subsection*{Appendix C: Perturbation theory for $\rho\!\ll\!1$}
\setcounter{equation}{0}
\renewcommand{\theequation}{C.\arabic{equation}}
This Appendix gives a brief summary of the perturbation expansion of
Eq.(\ref{dt.Gn}).
For small values of the fluctuation parameter $\rho\!=\!\d^2/(2\s_e^2)$ a
first-order expansion yields
\be\label{Gn.appr}
G_n(\e,t|\e_0)\!\simeq\!G(\e,t|\e_0)
\!-\!n\rho\int_0^t\!d\t\int\!d\e'\int\!d\e''G(\e,t\!-\!\t|\e')\k(\e'|\e'')
G(\e'',\t|\e_0)
\ee
and thus, using $\int\!d\e G(\e,t|\e_0)\!=\!1$
\be\label{C.e.t}
\int\!d\e G_n(\e,t|\e_0)
\simeq \exp{\left[-n\rho\int_0^t\!d\t\!\int\!d\e'\k(\e')G(\e',\t|\e_0)\right]}
\ee
Here, $\exp{(-x)}\!\simeq\!(1\!-\!x)$ and the definition of
$\k(\e)$, cf. Eq.(\ref{ME.e}), has been used.
Inserting Eq.(\ref{C.e.t}) in Eq.(\ref{P.om.marg}) and using
Eq.(\ref{P.om.e}) shows that the resulting sum over $n$ can be
evaluated\cite{risken}.

In the short time limit, one can replace $G(\e',\t|\e_0)$ by its initial
value, $G(\e',\t\!=\!0|\e_0)\!=\!\d(\e'-\e_0)$, yielding
$\int\!d\e G_n(\e,t|\e_0)\!\simeq\!\exp{[-n\rho\k(\e_0)t])}$.
Thus, one finds Eq.(\ref{P.ic}) with $\l\!=\!\rho\k(\e)$ and
$p(\l)\!=\!p^{eq}(\e)$.

In the long time limit, on the other hand, one can write
$G(\e',\t\!=\!\infty|\e_0)\!=\!p^{eq}(\e')$ in Eq.(\ref{C.e.t}).
This gives $\int\!d\e G_n(\e,t|\e_0)\!\simeq\!\exp{[-n\rho\lg\k\rg t]}$, where
the mean relaxation rate is defined as $\lg\k\rg\!=\!\int\!d\e\k(\e)$.
Again, one recovers Eq.(\ref{P.ic}) with $p(\l)\!=\!\d(\l\!-\!\rho\lg\k\rg)$.
\end{appendix}
\section*{Figure captions}
\begin{description}
\item[Fig.1 : ] Sketch of the evolution of the optical emission spectrum.
The environmental relaxation gives rise to the Stokes shift of the observed
time-resolved emission spectra. Here $|g\rg$ and $|e\rg$ denote the ground
and excited electronic state, respectively. $I(\om,t)$ is the time-dependent
optical lineshape evolving from the initial absorption spectrum $S_g(\om)$
at $t\!=\!0$ to the steady state emission spectrum $S_e(\om)$ at long times.
\item[Fig.2 : ] $C(t)$ (upper panel) and $C(2t)\!-\!C(t)^2$ (lower panel)
versus scaled time $t/\t_K$ for $\rho=0.1$.
The parameters are chosen such that the $C(t)$ are well fitted to
$C(t)=\exp{(-(t/\t_K)^{\b_K})}$.
Dashed lines: $\b_K\!=\!0.65$, $\k_\infty\!=\!2.3\!\times\!10^3s^{-1}$,
$\s(\e)\!=\!3.055$;
full lines: $\b_K\!=\!0.50$, $\k_\infty\!=\!7.2\!\times\!10^4s^{-1}$,
$\s(\e)\!=\!3.65$;
dotted lines: $\b_K\!=\!0.35$, $\k_\infty\!=\!5.83\!\times\!10^9s^{-1}$,
$\s(\e)\!=\!5.03$.
Here, $\s(\e)$ is given in temperature units ($\b\!=\!1$).
Additionally, the values of $\e$ are restricted to the interval
$\e\in\{-\s(\e)^2,\s(\e)^2\}$, giving equilibrium probabilities
$p^{eq}(\e)\propto\eta(\e)e^{-\b \e}$ that are truncated Gaussians.
Note, that the prefactors include a factor $\exp{(-\b E_A)}$ and therefore
are only related in an indirect way to the attempt frequencies, typically on
the order of $10^{15}s^{-1}$.
\item[Fig.3 : ]
{\bf (a)}: $\xi(t)$ as defined in Eq.(\ref{sig.t.ng}) and
$C(2t)-C(t)^2$ versus scaled time for the same parameters as in Fig.2 for
$\b_K\!=\!0.5$. The difference between $\xi(t)$ and $C(2t)-C(t)^2$ shows the
deviations from the Gaussian approximation, Eq.(\ref{P.om}), for the
conditional probability.\\
{\bf (b)}: The quantities $\zeta_n(t)$, defined in Eq.(\ref{zeta.n}), versus
scaled time. In the Gaussian approximation the $\zeta_n(t)$ vanish identically,
$\zeta_n(t)\!=\!0$.
\item[Fig.4 : ] Double-sided Feynman diagram for the excitation and for the
emission of an electronic two-level system, cf. Eq.(\ref{St.allg}).
The solute-system initially is in thermal equilibrium in the ground state
($\varrho\!\propto|g\rg\lg g|$).
The interaction with the incoming radiation, the wavy lines, produces an
optical coherence ($|e\rg\lg g|$), which then is converted into an excited
state population ($|e\rg\lg e|$).
After photon emission, which again is accomponied with a coherence, the system
is again in a ground state population ($|g\rg\lg g|$).
\end{description}
\newpage
\end{document}